\documentclass{article}

 \usepackage[nonatbib, final]{neurips_2025_ml4ps}

\usepackage[utf8]{inputenc} 
\usepackage[T1]{fontenc}    
\usepackage{hyperref}       
\usepackage{url}            
\usepackage{booktabs}       
\usepackage{amsfonts}       
\usepackage{nicefrac}       
\usepackage{microtype}      
\usepackage{xcolor}         
\usepackage{amsmath}
\usepackage{tikz}
\usetikzlibrary{positioning, arrows.meta}
\usepackage{siunitx}	
\usepackage{float}
\usepackage{cleveref}
\usepackage{multirow}
\usepackage{printlen}
\usepackage{makecell}
\usepackage[noadjust]{cite}

\author{%
  Pranav Sampathkumar\\
  \And
  Tim Huege \\
  \And
  Andreas Haungs \\
  \AND 
  Ralph Engel\\
 Institute for Astroparticle Physics,\\
 Karlsruhe Institute of Technology,\\
  Karlsruhe, Germany
}

\newcommand{\vB}{v \times B}
\newcommand{\vvB}{v \times v \times B}
\newcommand{\xmax}{X_\mathrm{max}}

\title{Neural Network for Simulating Radio Emission from Extensive Air Showers}

\begin{document}

\maketitle

\begin{abstract}
Cosmic ray shower detection using large radio arrays has gained significant traction in recent years.
With massive improvements in signal modelling and microscopic simulations, the analysis of incoming events is still severely limited by the simulation cost of radio emission to interpret the data.
In this work, we show that a neural network can be used for simulating such radio pulses.
This work serves as a proof of concept that simple neural networks can be used for emergent deterministic macroscopic phenomena of microscopic simulations. 
We also demonstrate how such a neural network can be used for the physics use case of $\xmax$ reconstruction, while retaining comparable resolution to using full Monte-Carlo simulations for radio emission. Code available at \url{https://anonymous.4open.science/r/radio_nn-21BF/}.

\end{abstract}

\section{Introduction}
Cosmic rays from distant sources travel through vast distances to reach Earth. They then interact with the Earth's atmosphere to create a cascade of secondary emissions known as extensive air showers. 
The use of radio emission from extensive air showers (EAS) to study cosmic rays has gained significant traction in the past decade, with various advances to signal processing and instrumentation using massive arrays of antennas.
These advances have led to reliable reconstruction approaches for cosmic ray events using radio antennas.
These reconstructions provide valuable insights into the energy spectrum and mass composition of cosmic rays, while radio arrays are cost-efficient for deployment over vast areas.
Various current/planned experiments \cite{aera,lofar,SKA,icecube-gen2,rnog,grand} utilize radio emission in order to improve the study of cosmic rays. 
The current reconstruction methodology used for these arrays involves microscopic simulations of the electromagnetic emission from individual tracks of the particles.
This makes these simulations computationally heavy, as the individual contributions from all these tracks need to be calculated and added up for every antenna.
In some established analysis approaches \cite{aera_xmax_paper, lofar_xmax_paper}, multiple iterations of each measured event need to be simulated in order to perform a detailed study, while keeping track of the uncertainties arising in the radio measurements, which are noisy in nature.
This makes the computational cost a massive bottleneck in the reconstruction of new radio data at higher and higher energies. 

Previous works have been done in order to reduce this computational cost for big arrays of antennas using various interpolation methods \cite{my_icrc_2023,template_synthesis,pulse_interpolation}, but none generalize across event geometries and still rely on computations. In this work, we provide with a neural network which can simulate radio emission for a given position in a wide variety of event geometries.
We will then show that this neural network can be used for reconstruction via the shower maximum $X_{max}$, for individual events, thus reducing the simulation times drastically for estimating the primary mass from radio detection. Though the microscopic interactions in the shower are probabilistic, this work shows that the deterministic electromagnetic emission can be modelled with a simple neural network instead of the harder traditional problem of capturing the target distribution using generative methods.

\section{Training Data}
This work uses the vast library of simulations built for analysing events within AERA (Auger Engineering Radio Array) \cite{aera_xmax_paper}\footnote{We are grateful to the Pierre Auger Collaboration for allowing us to use their simulation set within this work.}.
The network is trained in AERA's frequency band, 30-80MHz and specifically with the atmosphere above AERA. 
These simulations were set up with the Monte-Carlo code CORSIKA 7.7100 \cite{corsika7} with radio emission from particle tracks done with the CoREAS extension \cite{coreas}.
The resulting library has 2158 different shower parameters with 27 different iterations of each of these showers, resulting in a total of around 58k simulations.
This work serves as a proof of concept within the context of AERA
\footnote{
However, the model can be fine-tuned with a smaller dataset for changes in experimental setup and frequency bands.  Such a model has been successfully trained for LOFAR simulations \cite{denis_lofar}.
}, and about how, by re-framing a problem to deterministic terms, the problem becomes simpler. 
We use the star-shaped antenna positions in each of these simulations (240 antennas). Of this dataset, 80\% of the simulations are used for training and the remaining 20\% are used for testing purposes. 
\section{Neural Network Model}
\label{sec:nn_model}
Neural networks have been gaining substantial interest as an alternative simulation tool for physics. 
The rise of generative AI techniques has helped speed up physics simulations significantly.
We use a fully connected neural network for our case here, which is being directly trained to generate the pulses at each antenna position. Previous work has been done to perform particle cascades using various generative models \cite{calochallenge2022,caloclouds2,calodiffusion,calodream,caloflow,caloforest,caloinn,caloman,calopointflow,caloscore,calovq}. High fidelity in high energy cascades is extremely hard and compute intensive. Here, we take the alternate hybrid approach, where the probabilistic particle cascade is done explicitly and captured using key physics parameters (Shower Maximum ($X_{max})$, Electromagnetic energy $E_{em}$ ) and the deterministic part is done using a neural network. 
This network uses the atmospheric depth at which the particle count reaches its maximum ($\xmax$), the electromagnetic energy, geomagnetic angle, density and height at $\xmax$, primary energy of the incoming cosmic ray, arrival direction of the shower via zenith and azimuth angles and the antenna position in shower coordinates as input. This is inspired by point cloud techniques \cite{caloclouds2}, where the simulation of the entire emission is done in a geometry-independent manner with a network modelling each antenna position.
We have a total of 11 input parameters, and the network predicts pulses in both polarisations.
The pulses are predicted with 256 time bins in each polarisation with a timing resolution of 1\,ns, resulting in 512 output nodes for the network.
The network has 8 hidden layers, and the schematic is given in \Cref{fig:nn_schematic}.
There are no skip connections, as it was found that the gradients do not vanish at this depth.
We scaled the input values by nominal values so that the distribution lies between -1 and 1. 
The training process has an L1 norm ($\mathcal{L} = | x- y|$) as the loss function, and we weight the second polarisation more, since the dominant emission is within a single polarisation.
Weighting the weaker polarisation accordingly makes sure that both polarisations are learnt correctly.
The L1 norm was chosen so that the weaker pulses, whose contributions to the loss will be low, are still represented sufficiently.
The minimisation algorithm used is ADAM \cite{adam}, and regularisation is done via weight decay during minimisation.
The pipeline is implemented using Pytorch and Numpy \cite{pytorch,numpy}.
The training procedure is fast and can be completed within a week on a local desktop computer on CPUs \footnote{Tested on an AMD Ryzen 7 PRO 3700 8-Core Processor, 3.6 GHz}.
After training, inference requires only a few hundred milliseconds for each antenna position simulated.
Neural networks also have the advantage of being optimised to use GPUs and multi-threaded CPUs. Thus, this helps parallelise the radio-emission simulations, which is an ongoing effort by the community \cite{c8_coreas_parallel}.
\begin{figure}[h]
    \centering
\begin{tikzpicture}[
    node distance=0.8cm and 0.5cm,
    every node/.style={draw, minimum height=1cm, minimum width=1.5cm, align=center},
    ->, >=Stealth
]

\node (input) {Input\\(11)};
\node (l1) [right=of input] {64};
\node (l2) [right=of l1] {128};
\node (l3) [right=of l2] {512};
\node (l4) [right=of l3] {512};
\node (l5) [below=of l2] {512};
\node (l6) [right=of l5] {1024};
\node (l7) [right=of l6] {1024};
\node (l8) [right=of l7] {1024};
\node (output) [right=of l8] {Output\\(512)};

\draw[->] (input) -- (l1);
\draw[->] (l1) -- (l2);
\draw[->] (l2) -- (l3);
\draw[->] (l3) -- (l4);
\draw[->] (l4.south) -- ++(0,-0.5) -- ++(-4.025,0) -- (l5.north);
\draw[->] (l5) -- (l6);
\draw[->] (l6) -- (l7);
\draw[->] (l7) -- (l8);
\draw[->] (l8) -- (output);

\end{tikzpicture}
    \caption{Neural network schematic for generating radio pulses:
    The 11 inputs are described in \Cref{sec:nn_model}, and the 512 outputs are the 256 time bins in the 2 polarisations.
    The network produces a pulse, given an antenna position and shower parameters.
    The model has about 4M parameters, and has a memory footprint of 19MB. 
    }
    \label{fig:nn_schematic}
\end{figure}
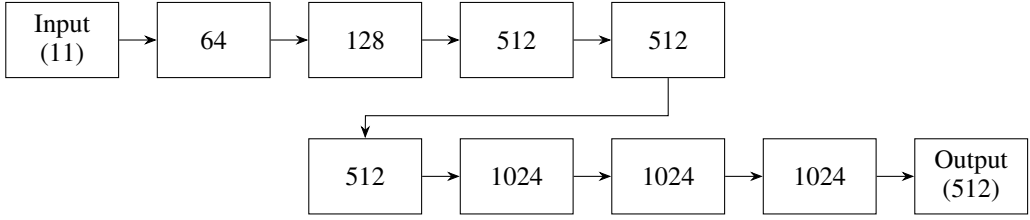
\begin{figure}
\centering
 \includegraphics[width=0.45\textwidth]{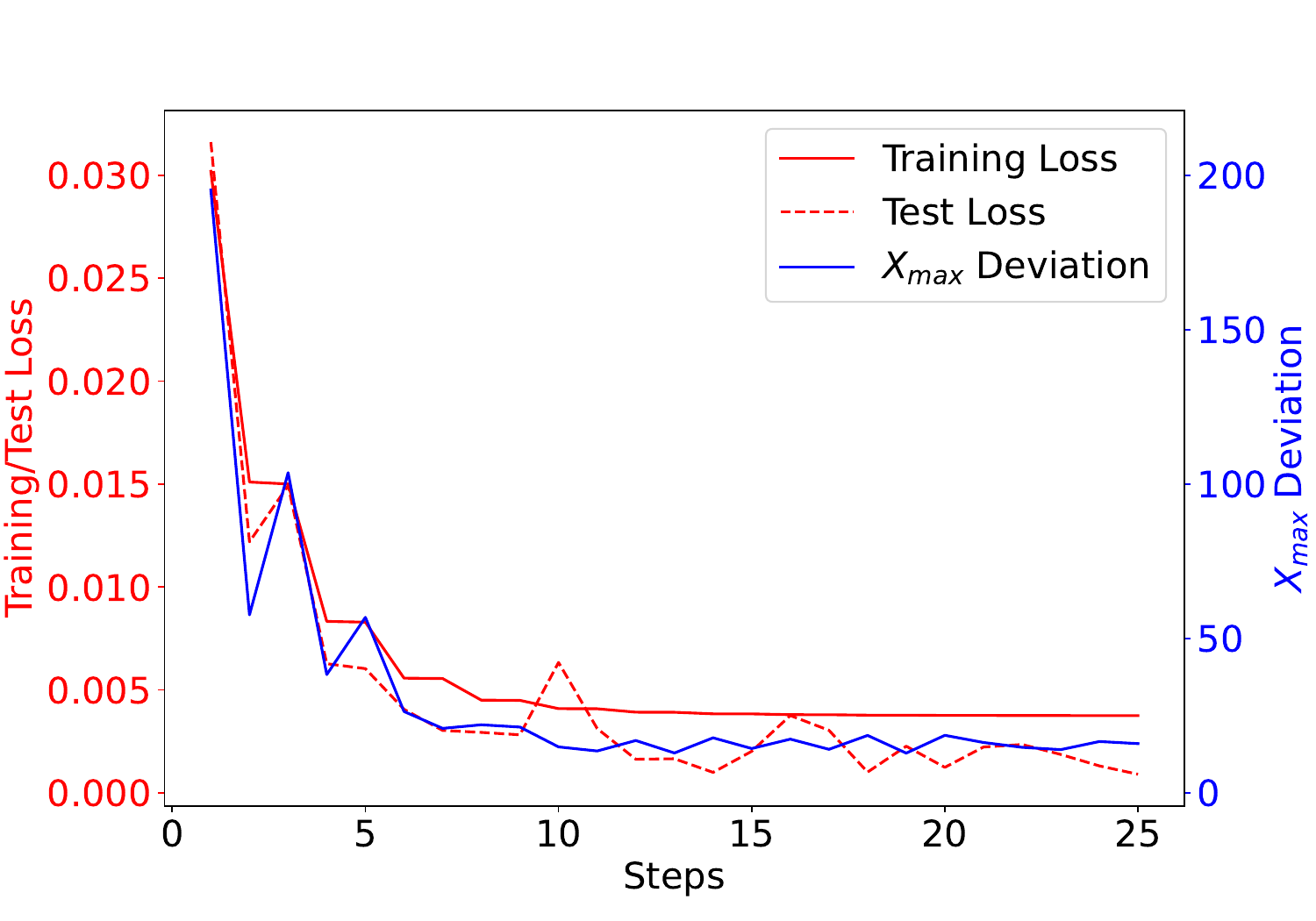}
 \includegraphics[width=0.45\textwidth]{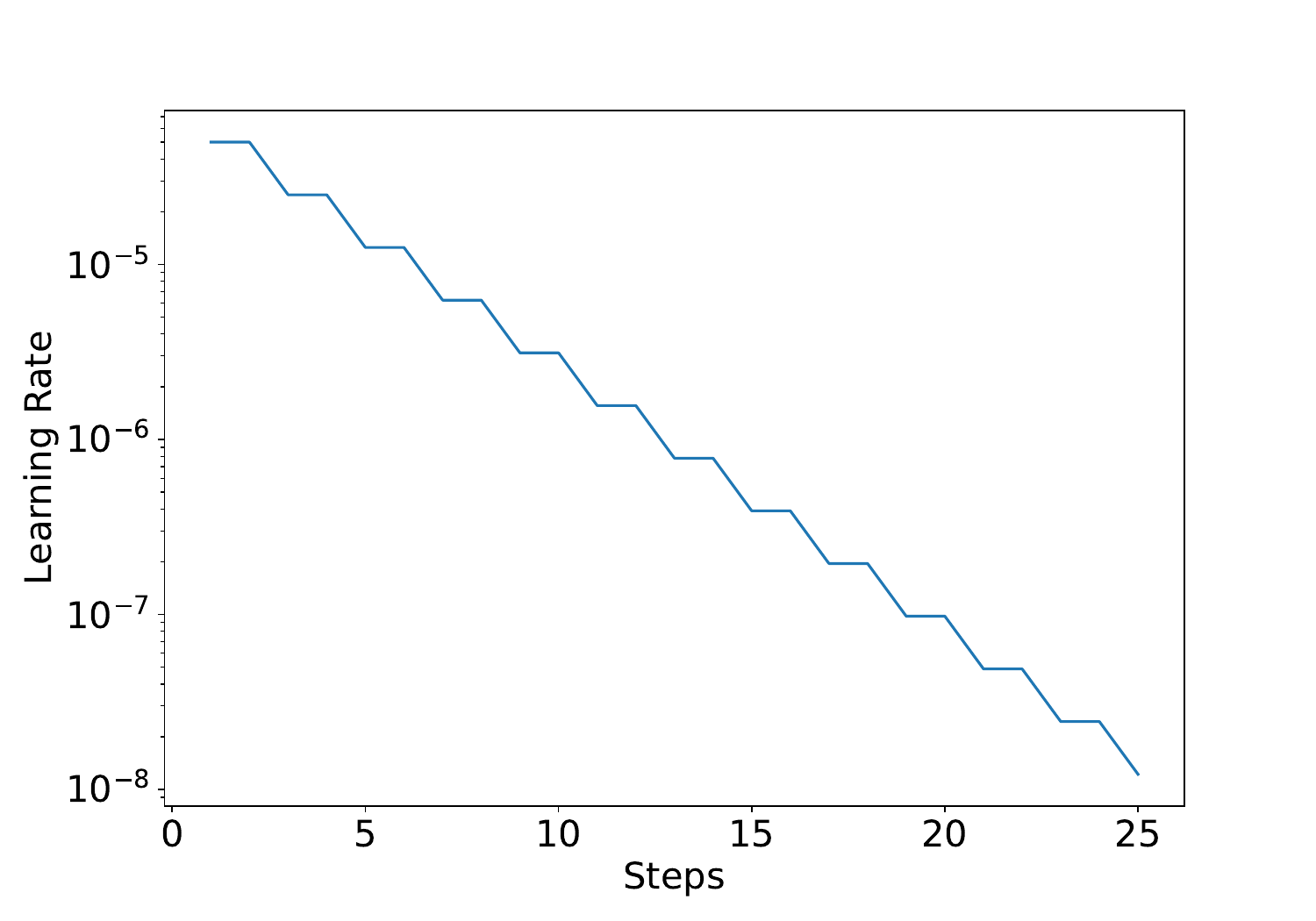}
\label{fig:loss_and_lr}
\caption[Training process for the neural network, showing loss, learning rate and $\xmax$ accuracy with epochs]{This figure on the left shows the change in loss functions as a function of training steps (or epochs). The figure on the right shows how the learning rate is varied with the training steps. In the plot on the left, the typical deviation from the true $\xmax$ for a $\xmax$ reconstruction procedure for a random shower is shown. }
\end{figure}
\section{Performance}
\subsection{Pulses}
\begin{figure*}[!htb]
    \centering
    \begin{tabular}{m{0.15\linewidth}|>{\centering}m{0.37\linewidth}|>{\centering\arraybackslash}m{0.37\linewidth}}
    & CoREAS & Neural Network \\
    \hline
    \makecell{$\vB$\\ polarization} & 
    \includegraphics[width=\linewidth]{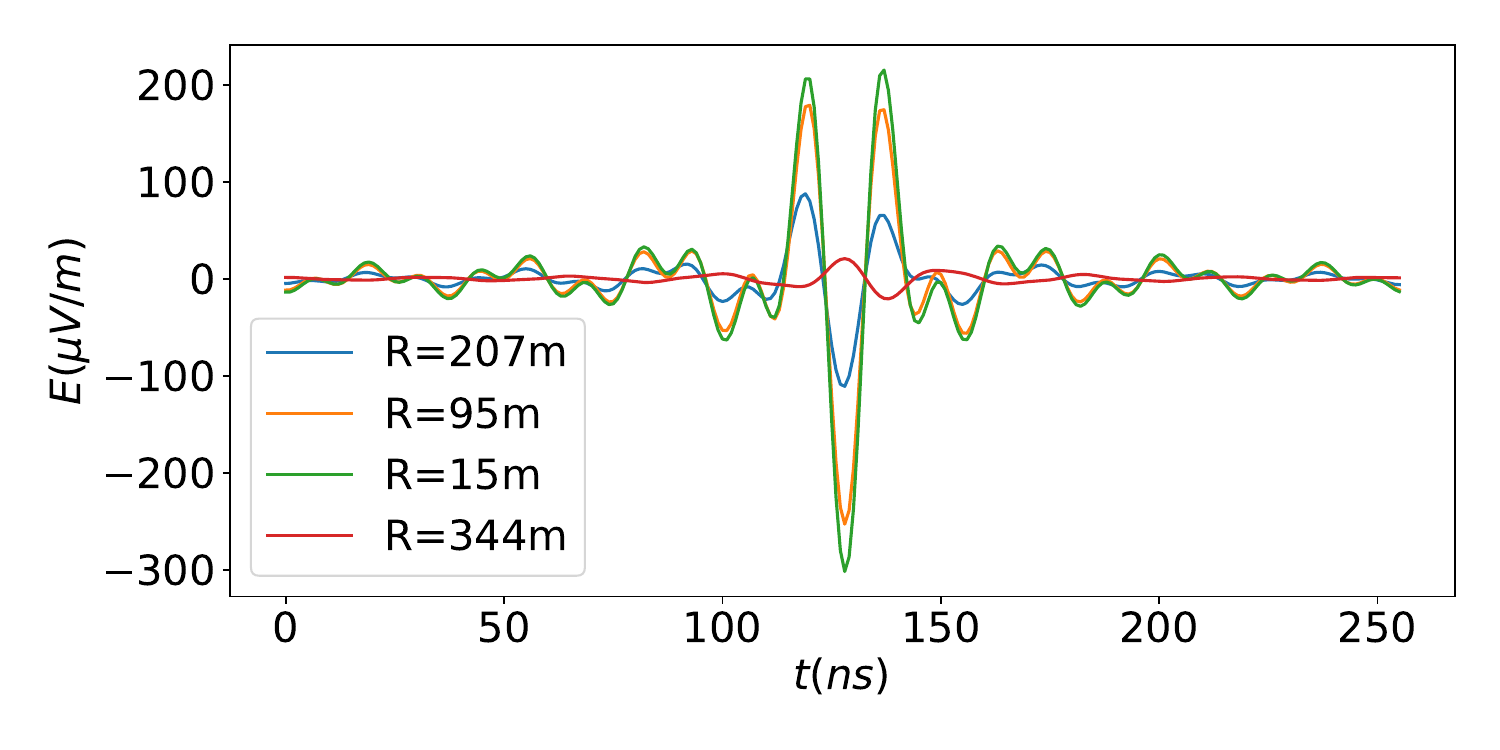} &
    \includegraphics[width=\linewidth]{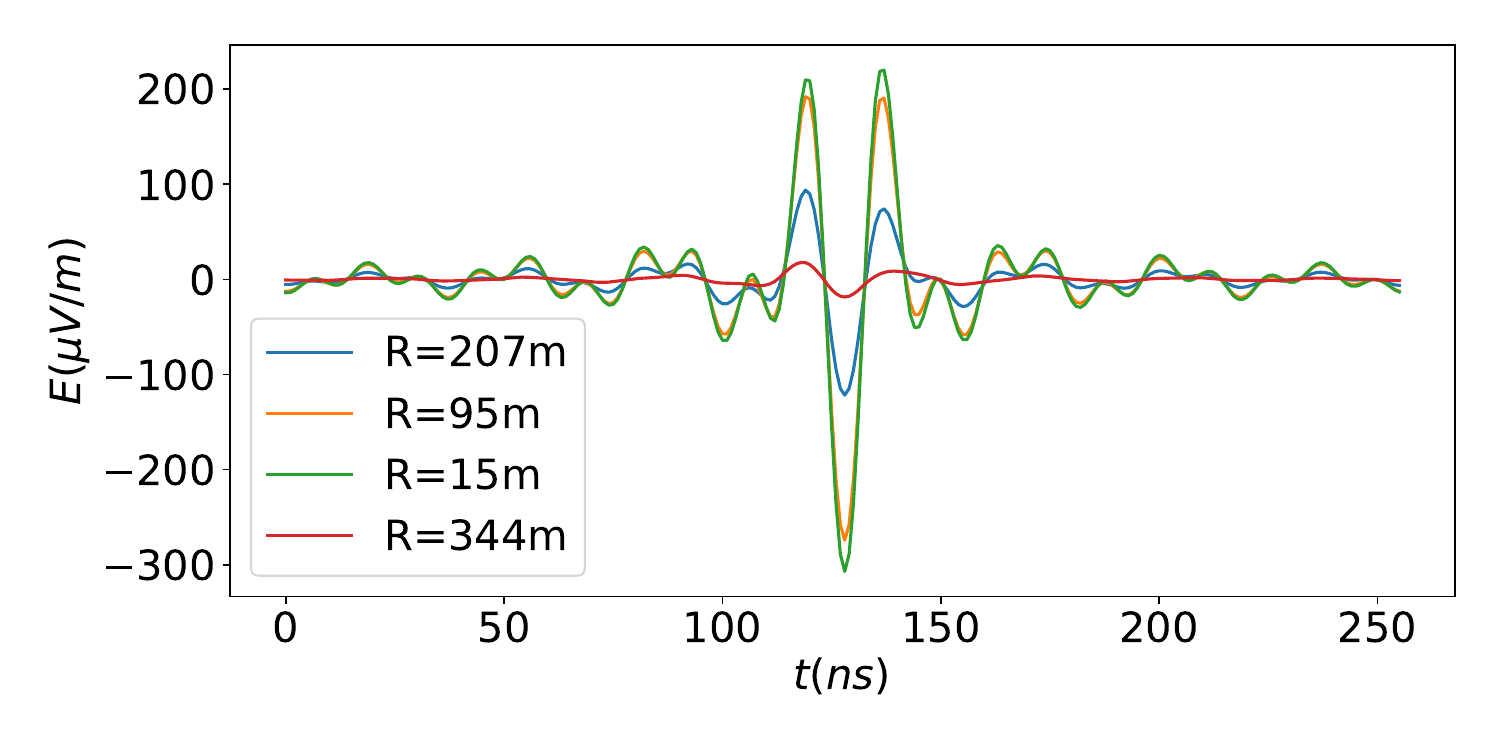}\\
    \hline
    \makecell{$\vvB$\\ polarization} & 
    \includegraphics[width=\linewidth]{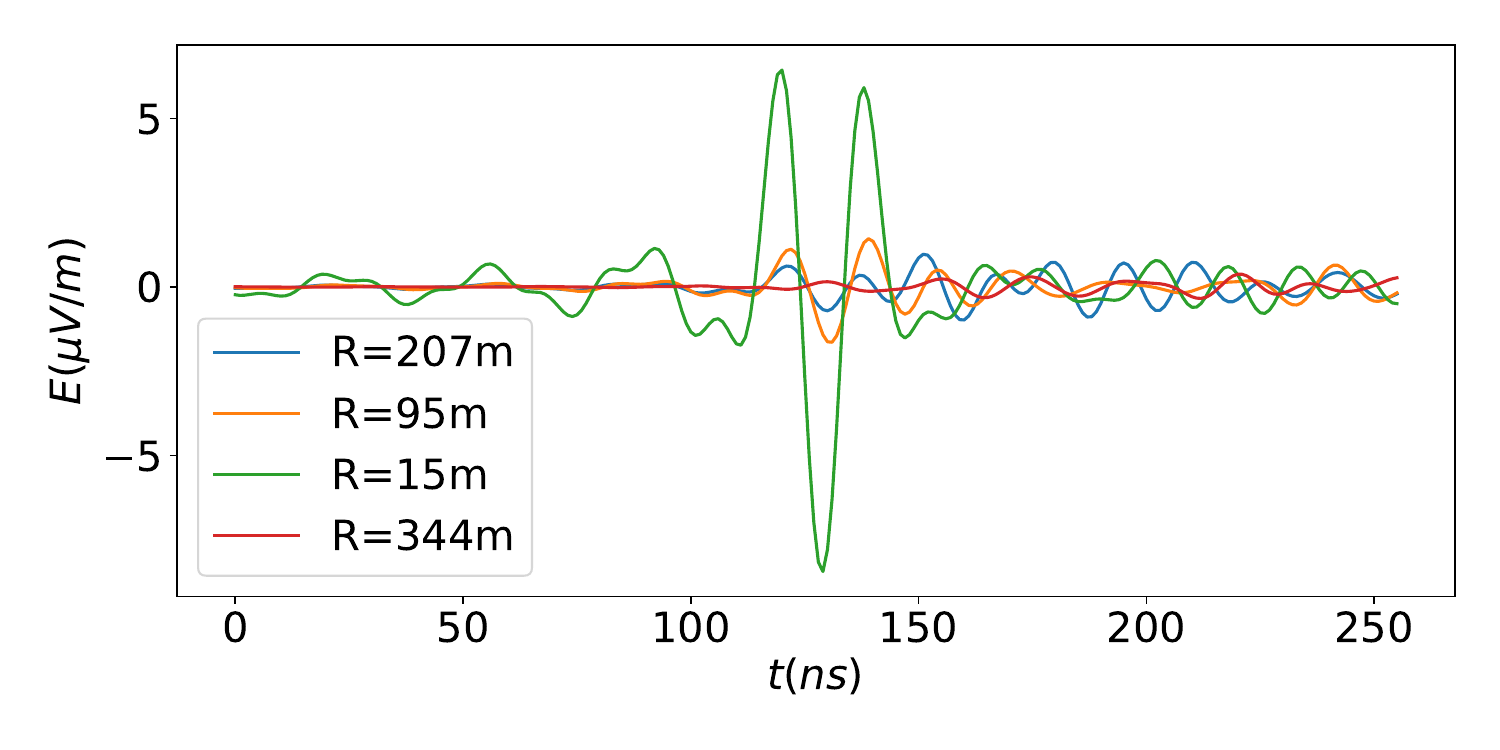} &
    \includegraphics[width=\linewidth]{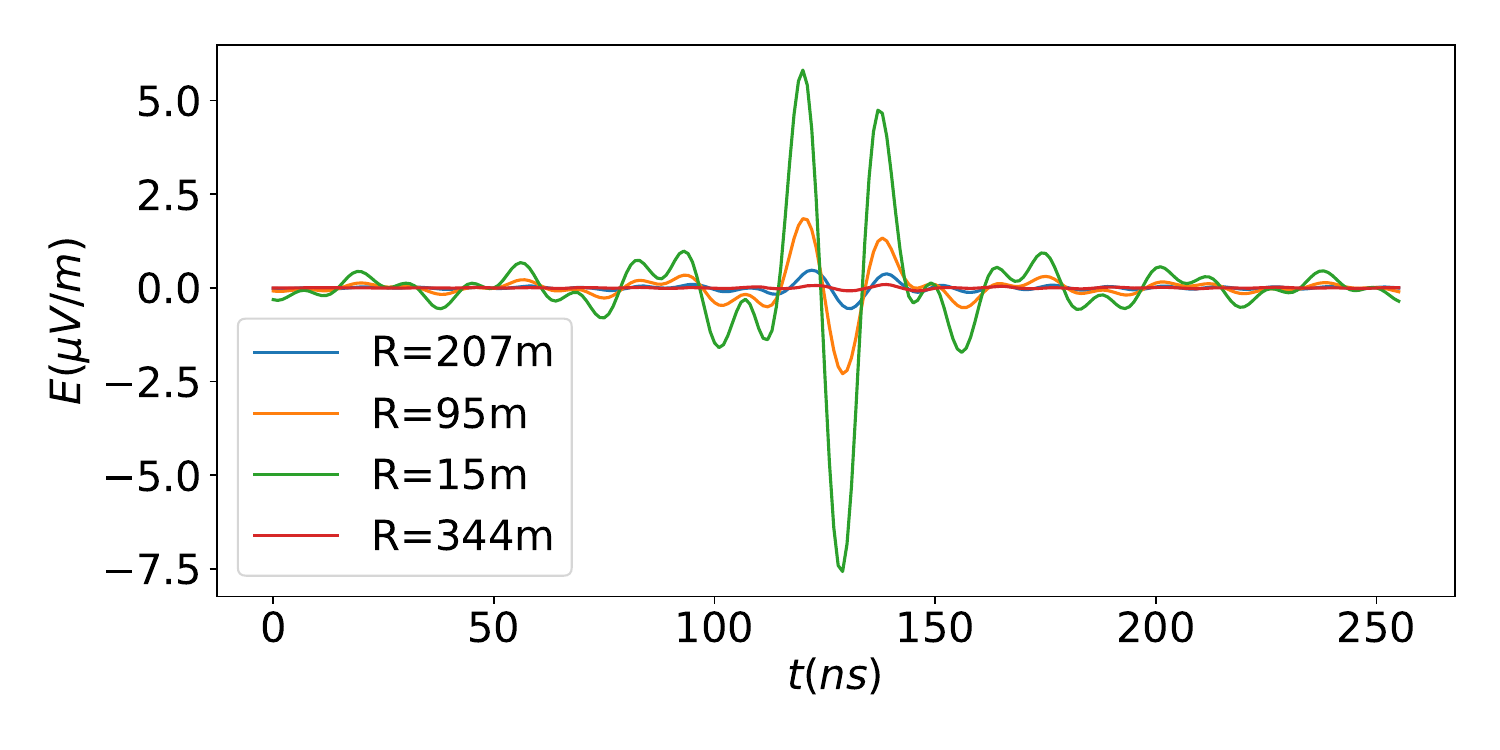}\\
    \end{tabular}
    \caption{Comparison of generated radio pulses between a CoREAS simulation and neural network-based simulation for both polarizations at antennas at various distances from the shower core (R). 
    This is a simulation of a single event.
    }
    \label{fig:pulses}
\end{figure*}

The neural network can directly produce centred pulses in the trained time resolution. We can see that the pulses agree qualitatively. We study the variation with pulses with varying geometries and $\xmax$ to see the agreement, of which one example is shown in \Cref{fig:pulses}. The energy deposited per unit area (the fluence) is critical in reconstructing shower parameters based on measured signals. The difference in fluence was studied, and the fluences are found to mostly match within 10\% accuracy. The normalised correlation between pulses was also studied, and we find that the pulses are highly correlated for strong pulses, making it very useful in real-world settings where weaker pulses are washed out by noise. Thus, the network shows promise and can be used for $\xmax$ reconstruction. 

The error associated with the energy fluence raises drastically in fluences less than \qty{1}{\electronvolt\per\square\metre}, raising up-to an $100\%$ error for fluences less than \qty{e-2}{\electronvolt\per\square\metre}. While, this can be improved with sufficient weighting of the weaker pulses, this is unnecessary, as the low fluence measurements in a realistic setting is completely dominated by ambient radio noise, and has too low of a signal-to-noise radio to be considered a reliable measurement. Thus, the reconstruction accuracy is reported with this simplistic model.

\subsection{$\xmax$ reconstruction}
In order to check the validity of the network for $\xmax$
reconstruction, we do a simplistic $\xmax$ reconstruction procedure which is inspired by the fluence-based method used by many experiments \cite{aera_xmax_paper, lofar_xmax_paper}.
These approaches work by calculating the fluence at each antenna and then comparing it with the fluence of various simulated events.
The $\chi^2$ difference between the measured and simulated fluences is calculated with varying $\xmax$, and following which the point of minimum $\chi^2$ is interpolated and taken as the reconstructed $\xmax$.

The $\chi^2$ difference between the measured shower and simulations is defined as
\begin{align}
    \label{eq:chi2}
    \chi^2 = \sum_\mathrm{antennas} \left( \frac{P_\mathrm{ant} - S^2 P_{sim}(x_\mathrm{ant}, y_\mathrm{ant})}{\sigma_\mathrm{ant}} \right)^2
\end{align}
where S is a scaling factor which accounts for multiplicative deviations between fluence measurements and simulations. $\sigma_{ant}$ denotes the uncertainties in fluence measurements at every antenna. For our purposes, since we are comparing simulations with simulations, we set $S=1$  and assume a particular model, described below, for the uncertainties. 

For the noise model, we assume a Gaussian noise (representing antenna-to-antenna gain variations). The method then involves freely simulating around the initial estimate of $\xmax$. A parabola is then fit to the $\chi^2$ values and the minimum of the parabola is used for the reconstructed $\xmax$ value.
Since the reconstruction method involves limited sampling, the parabola fit becomes biased with no equal number of simulations on either side of the parabola, leading to a bias in lower $\xmax$.

\subsection{Bias and Resolution of $\xmax$ reconstruction}
To benchmark the performance of using the neural network as a replacement for per-event CoREAS simulations in such an $\xmax$ reconstruction approach, we study the bias and resolution of the $\xmax$ reconstruction procedure for varying noise levels.
We see that both the bias and resolution for the simulations are similar between neural network predictions and per-event CoREAS simulations, making neural networks a viable simulation tool for simulating radio pulses. Even the variation in bias and resolution with respect to various shower geometries was found to be consistent.
The total bias and resolution for all the events, irrespective of $\xmax$, are given in \Cref{tab:bias_res_results_gaussian_noise}. 
The neural network can be further optimized to improve these, but that is left for the future, since the current model's purposes can be broader than this $\xmax$ reconstruction. 
\begin{table}[h]
    \centering
    \caption{The total bias and resolution for the $\xmax$ reconstruction procedure (in units of \qty{}{\gram\per\square\centi\metre}).}
    \begin{tabular}{ccccc}
         &CoREAS - Bias& CoREAS - Res. & NN - Bias & NN - Res.\\
        \hline
         5\% Noise &  - 12.87 & 32.22 & -12.71 & 33.13\\
         10\% Noise &  - 10.36 & 31.42 & -11.79 & 32.38\\
    \end{tabular}

    \label{tab:bias_res_results_gaussian_noise}
\end{table}

The reported difference between the two major radio emission models, CoREAS and ZHS is reported to be around 10\% for the total radiation energy \cite{coreas_zhs_comp}. Additionally, the energy fluences are also in the same range of $\sim 10\%$, which is comparable to shower to shower fluctuations in the emission, and well within the systematic uncertainties of the detectors in the \qtyrange{30}{80}{\mega\hertz} range. The effect on the emission model for $\xmax$ reconstruction is a variation of $\sim$\qty{11}{\gram\per\square\centi\metre}. Thus the differences in the neural network reported in this work, is well within the range of uncertinities due to the emission model, and can thus be used for $\xmax$ reconstruction pipelines reliably. 

The trained model can also be used to fine-tune for other experiments to generate a model with smaller datasets. Further work was done in our group, to extend the model by fine-tuning it for a different atmosphere in a different experimental step. This, fine-tuning was done with a much smaller dataset and was successfully tested for reconstruction too. Thus this model also serves as a valuable starting point in the astroparticle physics community, being trained on an experiment with the most available data, lending itself to be fine-tuned for other experiments without the same computational load for dataset generation. 
\section{Conclusion}
In this work, we presented a neural network that can be used to perform simulations of radio emissions from extensive air showers.
It provides a massive performance boost, as pulse predictions using neural networks can be performed almost instantly compared to traditional simulations, which sometimes take weeks.
Additionally, we also saw that the $\xmax$ reconstruction performance of the neural network is on par with using the CoREAS simulations.
The network is also shallow and has a low memory footprint.
Also, neural networks provide the advantage of being differentiable, and thus can be effectively used with other frameworks such as Information Field Theory 
\cite{keito_ift}.
While this current work is a proof of concept, the network can be extended with additional input parameters, such as the detailed longitudinal profile of the entire Extensive Air Shower and thus can be used along with other frameworks to reconstruct the longitudinal profile from measured signals.
Thus, the developed setup and the model can serve as a valuable contribution to the study of radio emission and the detection of cosmic rays using radio arrays, and also serves to show how re-framing a problem drastically simplifies it to the broader community using generative AI in physics.

\bibliographystyle{JHEP}
\bibliography{references}


\appendix
\end{document}